\documentstyle[12pt]{article}
\renewcommand{\theequation}{{\rm
\thesection.\arabic{equation}}}
\newcommand{\be}{\begin{equation}}
\newcommand{\ee}{\end{equation}}
\newcommand{\bea}{\begin{eqnarray}}
\newcommand{\eea}{\end{eqnarray}}
\newcommand{\nn}{\nonumber}

\newcommand{\ci}{{\rm i}}

\newcommand{\ket}[1]{{|#1\rangle}}
\newcommand{\bra}[1]{{\langle#1|}}
\newcommand{\om}[1]{{{\omega_{#1}}}}

\newcommand{\threej}[6]{\left(\begin{array}
{ccc}#1&#2&#3\\#4&#5&#6\end{array}\right)}

\begin{document}
\baselineskip 3.4ex
\begin{center}
{\huge \bf  Dynamical Stability and
Quantum Chaos of Ions in a Linear Trap}
\bigskip\bigskip
\bigskip

{\large G.P. Berman} [1] \footnote{Address correspondence to
Dr. G.P. Berman, Theoretical Division T-13,  MS B-213,  Los Alamos
National Laboratory, Los Alamos NM 87545, USA.  Email
gpb@lanl.gov.},
{\large D.F.V. James} [2],
{\large R.J. Hughes} [3],\\
{\large M.S. Gulley} [4],
{\large M.H. Holzscheiter} [3]\\
\bigskip
{\small [1] Group T-13 and CNLS, [2] Group T-4, [3] Group P-23, [4] Group P-25} \\
{\small University of California, Los Alamos National Laboratory}\\
{\small Los Alamos, NM87545, USA}\\
\bigskip
{\small and }\\
{\large G.V. L\'opez}\\
\bigskip
{\small Departamento de F\'isica, Universidad de Guadalajara} \\
{\small Corregidora 500, S.R. 44420, Guadalajara, Jalisco, Mexico}\\
\bigskip
{to be submitted to {\em Physical Review A}}\\
\bigskip
\bigskip
{\large \bf Abstract}
\end{center}
The realization of a paradigm chaotic system, namely the harmonically
driven oscillator, in the quantum domain using cold trapped ions
driven
by lasers is theoretically investigated. The simplest characteristics
of regular and chaotic dynamics are calculated. The possibilities of
experimental realization are discussed.
\bigskip
\begin{center}
PACS numbers:42.50.Vk, 05.45.Mt, 32.80.Pj\\
LA-UR-99-1217\\
\end{center}


\newpage
\section{Introduction}
One of the major difficulties in developing quantum technologies,
such as
quantum computers \cite{fortsch,bdmt}, are the different kinds of
specifically
quantum dynamical instabilities that can occur due to interactions
between different degrees of freedom and resonant interaction  with
the external fields.
These instabilities differ from dynamical instabilities in classical
systems, which are
usually connected
with strong dependence of trajectories on the initial conditions and
on the values of parameters.  Small variations of initial conditions
or parameters lead to large
deviations in time of the corresponding trajectories. If the speed of
this deviation is exponential, the system becomes chaotic, and the
appropriate methods of
description are statistical rather than deterministic. But for
quantum systems, the notion of a trajectory is not well defined. This
is one of the main reasons why most of the well-developed methods for
stability analysis can not be directly applied to quantum systems.
Moreover, as was first shown theoretically by Berman and Zaslavsky
\cite{ber1,ber1.1} (see also \cite{ber2}), even in a ``deep''
quasiclassical region, classically chaotic systems can have quantum
dynamics that is very different from the corresponding classical
dynamics.

Another important phenomenon which takes place in quantum systems which
are classically chaotic is {\em quantum nonlinear resonance} (QNR),
which was first introduced and investigated theoretically by Berman
and Zaslavsky \cite{ber3}. QNRs are quantum manifestation of
nonlinear resonances which play very important role in classically
chaotic systems \cite{chir1}-\cite{ll}. Interactions of QNRs are
known to be intimately connected to the transition to quantum chaos
\cite{ber4}-\cite{ber6}. In the simplest situations, QNRs occur when
a bound quantum system whose energy levels are not equally spaced is driven
by a
resonant perturbation. A QNR is characterized by two main parameters:
the number of quasi-energy levels, $\delta n$, which are ``trapped''
in the potential well of the resonance, and the characteristic
frequency of slow phase oscillations, $\Omega_{ph}$.  Isolated QNRs
imply stable quantum dynamics; overlapping QNRs cause a transition to
quantum chaos. QNRs are very general phenomena in non-integrable
quantum systems, and can be thought of as ``quasi-particles'' of
quantum chaos (for more details, see chapter 9 of Reichl's recent
book \cite{reic}, which is devoted to the transition to quantum chaos
caused by interaction of QNRs). Until now, QNR effects have been
experimentally investigated using Rydberg atoms in a resonant
micro-wave field \cite{bayf}. Understanding the instabilities connected with
overlapping QNRs is important for fundamental
problems related to the transition to quantum chaos, and for
the design of experimental devices (such as quantum computers based on ion
traps) in which these instabilities may cause significant problems.
To study the characteristic parameters of both isolated QNR's and the
problems related to interaction of QNRs it is important to choose a
model which (a) involves regulated (and relatively small) number of
interacting QNRs; (b) can be implemented experimentally in quantum
and quasiclassical regions of parameters.

In this paper we introduce a quantum model which is convenient for
investigation  of quantum dynamical instabilities and transition to
quantum chaos based on overlapping of QNRs. The model consists of a
single ion confined in a radio-frequency Paul trap and which interacts
with a resonant laser field. In the classical limit, this model
reduces to the well-known model of a
linear oscillator interacting with a plane electromagnetic wave,
and was investigated in \cite{zas2} (see also references therein).
The main advantage of our model is that the number of interacting
QNRs can be regulated, for example,  by varying the intensity of the
laser radiation, which is difficult to achieve in other
models based on the kicked interaction
\cite{ber7}-\cite{raizen2}.

Devices based on trapped ions have been used to investigate
experimentally fundamental aspects of quantum mechanics
\cite{mmkw,wmmklibbbm}, as well as for important technological
applications such as optical frequency standards \cite{freqstan} and
quantum computing \cite{fortsch, qcomp}. Ions are confined by a combination of
a rotating quadrupole potential (induced by the rod electrodes) and a
weak electrostatic potential (induced by the conical endcap
electrodes).  The ions, once trapped, can be cooled by standard
Doppler cooling and by a optical pumping method (``sideband
cooling''), which can cool multiple
ions down to the quantum mechanical ground state of the trapping
potential. In an ion trap quantum computer, information can be stored in the
internal quantum states of the ions (which constitute the quantum
bits, or
``qubits'' of the computer), and, using ultra narrow bandwidth lasers,
quantum gate operations can be realized between pairs of
qubits using quantum states of the collective motion of
the ions in the harmonic confining potential as a quantum
information bus \cite{dfvj}.
As such devices
are specifically  designed to investigate experimentally
the preparation, evolution and measurement of quantum systems
with large dimension Hilbert spaces, the linear ion trap
is an ideal apparatus to investigate
the problems of quantum dynamical stability, the transition to
quantum chaos, and the spectroscopy of quantum nonlinear resonances.
In this paper, we present the main elements of the derivation of our
model --  a quantum linear oscillator driven by a monochromatic wave,
 and the
 preliminary analytical and numerical results on the
 classical and quantum dynamics in different regions of parameters.

The paper is organized as follows.  In section 2, the
theory of how a trapped ion can be driven by laser fields
in the manner of the harmonically driven
oscillator is described in detail.  It should be
noted in particular how similar the arrangement and laser
requirements are to those employed in ion trap quantum
computer experiments.  In section 3, the {\em classical}
theory of the harmonically driven oscillator is discussed;
the quantum theory is described in section 4.  The connection
of this system with the solid-state Anderson localization model is
described in
section 5. The results of numerical simulations are
presented in section 6.  We conclude this paper with a
brief discussion of the possibilities for experimental verifications.

\section{Raman interactions of lasers and trapped ions}
\setcounter{equation}{0}
A single ion confined in a linear radio-frequency (rf)
trap may be described by an effective Hamiltonian
given by the formula \cite{dfvj}
\be
\hat{H}=\frac{1}{2m}\hat{p}^{2}+\frac{1}{2}m\omega^{2}\hat{x}^{2}+
\hat{H}_{I} ,
\ee
where $m$ is the mass of the ion, $\hat{x}$ ($\hat{p}$) is the
position (momentum) operator for the ion and $\omega$ is the
angular trapping frequency.  We are
only considering motion of the ion along one direction, namely
the axis of weak confinement of the trap; the ion is strongly
confined along the other two directions, transverse to the axis
and so we will assume that the motion in those directions can
be neglected (see Fig.1).

We will employ the interference of two laser beams acting
on the ion to realize experimentally our desired interaction
Hamiltonian $\hat{H}_{I}$ .  Such Raman interactions between lasers and
ions are a standard technique, and are described in
detail elsewhere \cite{GWJ}.  The ion, confined in
the harmonic trapping potential, will have many quantum levels
associated with both the internal (atomic) variables and
the external (motional) degrees of freedom.  We will confine
our attention two manifolds of such states, separated in
energy by an appreciable amount (see Fig.2).  What we have
in mind is a lower manifold consisting of the magnetic
sublevels of our ion, each level having a series of sidebands
associated with excitation of quanta of the external harmonic
oscillations; the upper manifold would then be the sublevels
of an excited state of the ion, with similar sidebands.  The
lasers with which the ion is interacting will be detuned
from the optical transition between the upper and lower levels,
so that there is a negligible probability of any of these
levels becoming excited: the lasers only cause redistribution
of population amongst the lower manifold of levels.  The upper
levels may then be adiabatically eliminated from the problem, and
one can therefore show that the matrix elements of
the effective interaction Hamiltonian for the
lower manifold is given by the formula \cite{GWJ}
\be
\bra{M}\hat{H}_{I}\ket{N}=-\sum_{L}\sum_{\alpha,\alpha'}
\frac{\Omega^{(\alpha)}_{ML}\Omega^{(\alpha')\ast}_{NL}}
{4\hbar (\om{L}-\om{N}-\om{\alpha'})}
\exp\left[\ci\left(\om{\alpha}-\om{\alpha'}\right)t\right] ,
\label{hone}
\ee
where the sum involving  $u$ is over all of the upper manifold levels
and the two sums involving $\alpha$ and $\alpha'$ are over
all of the applied laser fields, the Rabi frequency of
the $\alpha$-th laser being defined by
\be
\hbar\Omega^{(\alpha)}_{ML}=
\bra{M}\hat{d_{i}}
E^{(\alpha)}_{i}(\hat{\bf{r}}) \ket{L} .
\label{omegaone}
\ee

In Eq. (\ref{omegaone}), $\hat{d_{i}}$ is the $i$-th component
of the dipole moment operator ( $i=(1,2,3)$, standing for
the three Cartesian components of a vector, and summation over
repeated
indices being implied), $E^{(\alpha)}_{i}$ is the  $i$-th component
of the electric field from the $\alpha$-th laser (which is a function
of the ion's position operator, $\hat{\bf{r}}$), $\hbar \om{M}$ is
the energy of the $M$-th lower manifold level, $\hbar \om{L}$ is
the energy of the $L$-th upper manifold level, and $\om{\alpha}$ is
the angular frequency of the  $\alpha$-th laser.

To proceed, we will make the distinction between internal
and external degrees of freedom.  We can form a basis set
for the Hilbert space from a tensor product of a set of
internal quantum levels with a basis set for the external
degrees of freedom (for example the Fock states of the
Harmonic oscillator).  The set of internal states will be
divided between the upper and lower manifolds.  Thus we will
make the following substitution:
\bea
\ket{L}&\rightarrow&\ket{\lambda}\ket{l}\\
\ket{M}&\rightarrow&\ket{\mu}\ket{m}\\
\ket{N}&\rightarrow&\ket{\nu}\ket{n},
\eea
where $\ket{m},\ket{n}$ and $\ket{l}$ are members
of the basis states for the motion degrees of freedom,
$\ket{\lambda}$ is a member of the upper internal
manifold and $\ket{\mu}$ and $\ket{\nu}$ are members
of the lower internal manifold.  In this notation,
the matrix elements of
the Hamiltonian Eq. (\ref{hone}) become:
\bea
\bra{\mu}\bra{m}\hat{H}_{I}\ket{n}\ket{\nu}&=&-
\sum_{\lambda}\sum_{l}\sum_{\alpha,\alpha'}
\frac{\bra{\mu}\hat{d_{i}}\ket{\lambda}
\bra{\lambda}\hat{d_{j}}\ket{\nu}}
{4\hbar(\om{\lambda}-\om{\nu}-\om{\alpha'}+\om{l}-\om{n})}\nn\times \\
&&\bra{m} E^{(\alpha)}_{i}(\hat{\bf{r}}) \ket{l}
\bra{l} E^{(\alpha')\ast}_{j}(\hat{\bf{r}}) \ket{n}
\exp\left[\ci\left(\om{\alpha}-\om{\alpha'}\right)t\right] .
\label{htwo}
\eea

The average detuning of the $\lambda$-th upper
manifold level is be defined to be
\be
\Delta_{\lambda}=\om{\lambda}-\bar{\omega}_{\nu}-\bar{\omega}_{\alpha},
\ee
where $\hbar \bar{\omega}_{\nu}$ is the average energy of the
lower manifold and $\bar{\omega}_{\alpha}$ is the average
of the laser frequencies.  We will assume that, in the
denominator of the fraction appearing in Eq. (\ref{htwo}),
we can make the following approximation:
\be
\om{\lambda}-\om{\nu}-\om{\alpha'}+\om{l}-\om{n}\approx
\Delta_{\lambda}.
\ee
If we use the completeness property of
the external basis states (i.e.  $\sum_{l}\ket{l}\bra{l}=\hat{I}$,
where $\hat{I}$ is the identity operator), then we obtain the
following
formula for the Hamiltonian operator for the lower manifold states:
\be
\hat{H}_{I}=\sum_{\mu,\nu} \kappa_{\mu,\nu} \left(\hat{\bf{r}},t \right)
\ket{\mu}\bra{\nu},
\ee
where
\be
\kappa_{\mu,\nu} \left(\hat{\bf{r}},t \right)=
-\sum_{\lambda}
\frac{
\bra{\mu}\hat{d_{i}}\ket{\lambda}
\bra{\lambda}\hat{d_{j}}\ket{\nu}}
{4\hbar \Delta_{\lambda}}
E_{i}\left( \hat{\bf{r}},t \right)
E^{\ast}_{j}\left( \hat{\bf{r}},t \right),
\label{hthree}
\ee
and the total laser field $E_{i}$ is the sum of
the different laser components:
\be
E_{i}\left(\hat{\bf{r}},t \right)=
\sum_{\alpha}E^{(\alpha)}_{i}(\hat{\bf{r}})
\exp\left(\ci\om{\alpha}t\right).
\ee

\subsection{Two level systems}
Let us now assume that there are only two internal levels in
the lower manifold, which we will denote
$\ket{1}$ and $\ket{2}$.  As will be discussed below, this is a
reasonable assumption to make for the atomic systems we have in
mind.  Also, we introduce a special coordinate system: the two
internal levels are split by a magnetic field acting along
the $Z$ axis, which is the axis of quantization for the
internal levels of our ion.  The other two axes are the $X$
and $Y$ axes.  These axes do {\em not} necessarily coincide with
the $x,y,z$ directions introduced to describe the motion
of the ion in the trap.
In this case it is convenient to use the Pauli operators
for the system:
\bea
\hat{\sigma}_{1}&=&\ket{1}\bra{2}+\ket{2}\bra{1}, \\
\hat{\sigma}_{2}&=&\ci\left(\ket{1}\bra{2}-\ket{2}\bra{1}\right), \\
\hat{\sigma}_{3}&=&\ket{2}\bra{2}-\ket{1}\bra{1}.
\eea
Using these operators the Hamiltonian can be written
as follows:
\be
\hat{H}_{I}=h_{0}\left(\hat{\bf{r}},t\right) \hat{I}+
h_{i}\left(\hat{\bf{r}},t\right) \hat{\sigma}_{i},
\label{hfour}
\ee
where $\hat{I}$ is the identity operator $\left(\ket{1}\bra{1}
+\ket{2}\bra{2}\right)$ and
\bea
h_{0}\left(\hat{\bf{r}},t\right)&=&
\frac{1}{2}
\left[\kappa_{1,1}\left(\hat{\bf{r}},t\right)+\kappa_{2,2}\left(\hat{\bf{r}},t
\right)
\right], \label{hdefstart} \\
h_{1}\left(\hat{\bf{r}},t\right)&=&
\frac{1}{2}
\left[\kappa_{1,2}\left(\hat{\bf{r}},t\right)+\kappa_{2,1}\left(\hat{\bf{r}},t
\right)
\right],  \\
h_{2}\left(\hat{\bf{r}},t\right)&=&
\frac{1}{2\ci}
\left[\kappa_{1,2}\left(\hat{\bf{r}},t\right)-\kappa_{2,1}\left(\hat{\bf{r}},t
\right)
\right],  \\
h_{3}\left(\hat{\bf{r}},t\right)&=&
\frac{1}{2}
\left[\kappa_{2,2}\left(\hat{\bf{r}},t\right)-\kappa_{1,1}\left(\hat{\bf{r}},t
\right)
\right].
\label{hdefend}
\eea

For the special case that the lower manifold of internal
states consists of two magnetic sublevels of the
$^{2}S_{1/2}$ ground state of an alkali-like ion, and the
upper manifold is the two sublevels of the $^{2}P_{1/2}$
excited state whose Zeeman splitting is not too
big, the atomic matrix elements appearing in
Eq. (\ref{hthree}) can be calculated in closed form.  As a
result the components of $h$ appearing in Eq. (\ref{hfour})
reduce to the following simple form (see appendix A):
\bea
h_{0}\left(\hat{\bf{r}},t\right)&=&
-\chi \left|{\bf E}\left(\hat{\bf{r}},t\right)\right|^{2}\nn \\
h_{1}\left(\hat{\bf{r}},t\right)&=&
2\chi {\rm Im}
\left\{E_{Z}\left(\hat{\bf{r}},t\right)
E^{\ast}_{Y}\left(\hat{\bf{r}},t\right)\right\}\nn \\
h_{2}\left(\hat{\bf{r}},t\right)&=&
2\chi {\rm Im}
\left\{E_{X}\left(\hat{\bf{r}},t\right)
E^{\ast}_{Z}\left(\hat{\bf{r}},t\right)\right\}\nn \\
h_{3}\left(\hat{\bf{r}},t\right)&=&
2\chi {\rm Im}
\left\{E_{X}\left(\hat{\bf{r}},t\right)
E^{\ast}_{Y}\left(\hat{\bf{r}},t\right)\right\},
\label{aches}
\eea
where ${\rm Im} \left\{\ldots \right\}$ is the imaginary part of
the quantity in curly brackets and $\chi= A\pi \epsilon_{0}/4
k_{0}^{3}\Delta$
($k_{0}$ and $A$ being,
respectively, the wavenumber and the Einstein
A coefficient for the transition between the upper and lower
manifolds, $\Delta$ the laser detuning and $\epsilon_{0}$
the permittivity of free space).

The quantity proportional to $h_{0}\left(\hat{\bf{r}},t\right)$ in
Eq. (\ref{hfour}) represents a dynamical effect of the laser fields
on the ion which does not cause any effect on its internal degrees
of freedom; the term proportional to
$h_{3}\left(\hat{\bf{r}},t\right)$
represents a A.C. Stark shift of the two internals levels; the terms
proportional to $h_{1}\left(\hat{\bf{r}},t\right)$ and
$h_{2}\left(\hat{\bf{r}},t\right)$ represent transitions between the
two levels of the lower manifold.  If we make the requirement that
the lasers are plane-polarized along the axis of
quantization $Z$, then it is clear from the above formulas
that $h_{1}=h_{2}=h_{3}=0$ and only the first term involving
$h_{0}$ has any effect.

Let us assume that two laser beams, designated the pump (p)
and the Stokes (s) beams are present, both plane polarized in
the $Z$-direction, i.e.
\bea
E_{X}\left(\hat{\bf{r}},t\right)&=&0,\nn \\
E_{Y}\left(\hat{\bf{r}},t\right)&=&0,\nn \\
E_{Z}\left(\hat{\bf{r}},t\right)&=&
E^{(p)}\exp\left[-\ci\left({\bf k}_{p}\cdot\hat{{\bf
r}}-\om{p}t\right)\right] +
E^{(s)}\exp\left[-\ci\left({\bf k}_{s}\cdot\hat{{\bf
r}}-\om{s}t\right)\right].
\eea
The interaction Hamiltonian in this case is given by
\be
\hat{H}_{I}=\chi\left\{
\left|E^{(p)}\right|^{2}+\left|E^{(s)}\right|^{2}+
2\left|E^{(p)}E^{(s)\ast}\right|
\cos\left[\left({\bf k}_{p}-{\bf k}_{s}\right)\cdot\hat{{\bf
r}}-\left(\om{p}-\om{s}\right)t+\phi \right]
\right\},
\ee
where $\phi={\rm Arg}\left\{E^{(p)}E^{(s)\ast}\right\}$ is the
phase difference between the two lasers.
The constant terms involving $\left|E^{(p)}\right|^{2}$
and $\left|E^{(s)}\right|^{2}$ have no effect on the evolution,
and so will be neglected.  Thus the full Hamiltonian, including
the effect of the harmonic evolution of the ion along the weak
axis of the trap (but excluding the internal free evolution)
is
\be
\hat{H}={{\hat{p}^2}\over{2m}}+{{m\omega^2\hat{x}^2}\over{2}}+{{\varepsilon}
\over{k}}
\cos(k\hat{x}-\Omega t).
\label{hfin}
\ee
where $\Omega=\om{p}-\om{s}$.
The parameters $\varepsilon$ and $k$, which will feature prominently
in what follows, are given by
\bea
\varepsilon&=&
\frac{ A\pi \epsilon_{0}k}{2 k_{0}^{3}\Delta}
\left|E^{(p)}E^{(s)\ast}\right|,
\label{dimfuleps}\\
k&=&\left({\bf k}_{p}-{\bf k}_{s}\right)\cdot\bf{e}_{x}
\label{ohkay}
\eea
where $\bf{e}_{x}$
is the unit vector along the $x$-axis, i.e. the axis of weak
confinement in the trap.

\section{Classical Limit}
\setcounter{equation}{0}

In the classical limit $(\hat{p}\rightarrow p, \hat{x}\rightarrow
x)$, the Hamiltonian (\ref{hfin}) takes the form,
\be
H={{p^2}\over{2m}}+{{m\omega^2x^2}\over{2}}+{{\varepsilon}\over{k}}
\cos(kx-\Omega t),
\label{three.one}
\ee
The classical equations of motion in $(p,x)$ variables are
\be
\dot{p}=-{{\partial H}\over{\partial
x}}=-m\omega^2x+\varepsilon\sin(kx-
\Omega t), \label{pdot} \,\,\,
\dot{x}={{\partial H}\over{\partial p}}={{p}\over{m}} \label{xdot}.\\
\label{three.two}
\ee
Equations (\ref{three.two}) lead to the following
second order non-linear differential equation,
\be
\ddot{x}+\omega^2x={{\varepsilon}\over{m}}\sin(kx-\Omega t).
\label{three.three}
\ee

\subsection{Dynamics near resonances}
Assume that the driving frequency is close to a resonance, i.e.

\be
N\omega \approx \Omega ,
\label{three.four}
\ee
where $N$ is an integer. In this case, it is convenient to describe
a classical dynamics using the ``action-angle'' variables
\cite{zas2,zas3,zas4}, $(I,\varphi)$, which are related to the
variables
$(p,x)$ by the canonical transformation (see Appendix B),
\be
x=\sqrt{{{2NI}\over{\omega m}}}\cos\Bigg({{\varphi+\Omega
t}\over{N}}\Bigg),\quad p=-\sqrt{2NI\omega
m}\sin\Bigg({{\varphi+\Omega
t}\over{N}}\Bigg).
\label{three.five}
\ee
In the variables $(I,\varphi)$, the Hamiltonian (\ref{three.one})
takes the
form,
\be
H=(N\omega-\Omega)I+{{\varepsilon}\over{k}}\cos\Bigg(k\sqrt{{{2NI}
\over{\omega m}}}\cos\Phi-\Omega t\Bigg),
\label{three.six}
\ee
where
\be
\Phi={{\varphi+\Omega t}\over{N}} \nn.
\ee
The second term in (\ref{three.six}) can be expanded as a series
of Bessel functions $J_n(z)$,
\be
\cos\Bigg(k\sqrt{{{2NI}\over{\omega m}}}\cos\Phi-\Omega
t\Bigg)=\sum_{n=-\infty}^{\infty}J_n\Bigg(k\sqrt{{{2NI}\over{m\omega}}}
\Bigg)\cos[n(\Phi+\pi/2)-\Omega t].
\label{three.seven}
\ee
We have for the phase,
$n(\Phi+\pi/2)-
\Omega t$,
\be
n(\Phi+\pi/2)-\Omega
t=n{{\varphi}\over{N}}+{{n\pi}\over{2}}+{{n-N}\over{N}}\Omega t. \nn
\ee
Thus, the classical Hamiltonian (\ref{three.six}) can be represented
in the
form where the
unperturbed part and the perturbation are explicitly separated,
\be
H=H_0+H_{int},
\label{three.eight}
\ee
where
\bea
H_0&=&(N\omega-\Omega)I+{{\varepsilon}\over{k}}J_N\Bigg(k\sqrt{{{2NI}
\over{m\omega}}}\Bigg)
\cos(\varphi+N\pi/2),\label{three.nine}
\\
H_{int}&=&{{\varepsilon}\over{k}}\sum_{n\not=N}J_l\Bigg(k\sqrt{{{2NI}
\over{m\omega}}}\Bigg)
\cos\Bigg({{n}\over{N}}\varphi+{{n\pi}\over{2}}+{{n-N}\over{N}}\Omega
t\Bigg).
\label{three.ten}
\eea
In the $(I,\varphi)$ variables, the classical equations of motion are,
\bea
\dot I=-{{\partial H}\over{\partial \varphi}}&=&
{{\varepsilon}\over{k}}J_N(z)\sin(\varphi+\pi N/2)+\\
&&{{\varepsilon}\over{kN}}\sum_{n\not=N}nJ_n(z)\sin\Bigg({{n}\over{N}}
\varphi+
{{n\pi}\over{2}}+{{n-N}\over{N}}\Bigg). \label{three.eleven} \\
\dot \varphi={{\partial H}\over{\partial
I}}&=&N\omega-\Omega+\varepsilon\sqrt{{{N}\over{2m\omega
I}}}J_N^\prime(z)\cos(\varphi+\pi N/2)+\\
&&
\varepsilon\sqrt{{{N}\over{2m\omega
I}}}\sum_{n\not=N}J_l^\prime(z)\cos\Bigg({{n}\over{N}}\varphi+
{{n\pi}\over{2}}+{{n-N}\over{N}}\Bigg),
\label{three.twelve}
\eea
where,
\be
z=k\sqrt{{{2NI}\over{m\omega}}},
\label{three.thirteen}
\ee
is a dimensionless variable.
Equations (\ref{three.eleven}) and (\ref{three.twelve}) are
convenient when
analyzing the classical
dynamics in the vicinity of the resonance (\ref{three.four}). This
case
corresponds to small
values of $\varepsilon$,
\be
\varepsilon<\varepsilon_{cr}.
\label{three.fourteen}
\ee
Usually, the critical parameter $\varepsilon_c$ in (3.18) should be
found numerically (see
Sec. \ref{numres}).
Under the condition (\ref{three.fourteen}), classical dynamics can be
approximately
described by the Hamiltonian $H_0$ (\ref{three.nine}). The
corresponding
approximate
equations of motion follow from Eqs (\ref{three.eleven}) and
(\ref{three.twelve}),
\bea
\dot I&=&-{{\partial H}\over{\partial \varphi}}=
{{\varepsilon}\over{k}}J_N(z)\sin(\varphi+\pi
N/2),\label{three.fifteen}\\
\dot \varphi&=&{{\partial H}\over{\partial
I}}=N\omega-\Omega+\varepsilon\sqrt{{{N}\over{2m\omega
I}}}J_N^\prime(z)\cos(\varphi+\pi N/2).
\label{three.sixteen}
\eea
In the general case (for $\varepsilon$ large), it is more convenient to use
the following exact
equations written in $(I,\varphi)$ variables,
\bea
\dot
I&=&-{{\varepsilon}\over{N}}\sqrt{{{2NI}\over{m\omega}}}\sin\Bigg[k
\sqrt{{{2NI}\over{m\omega}}}
\cos\Bigg({{\varphi+\Omega t}\over{N}}\Bigg)-\Omega
t\Bigg]\sin\Bigg({{\varphi+\Omega
t}\over{N}}\Bigg),\label{three.seventeen}\\
\dot\varphi&=&N\omega-\Omega-\varepsilon\sqrt{{{N}\over{2m\omega I}}}
\sin\Bigg[k\sqrt{{{2NI}\over{m\omega}}}
\cos\Bigg({{\varphi+\Omega t}\over{N}}\Bigg)-\Omega
t\Bigg]\cos\Bigg({{\varphi+\Omega t}\over{N}}\Bigg).
\label{three.eighteen}
\eea

\subsection{Dimensionless variables}
To describe both the classical and quantum dynamics, it is convenient to
introduce
the following dimensionless variables,
\be
\left.
\begin{array}{cccc}
\displaystyle
\tau=\omega t,&
\xi=kx,&
\displaystyle\quad \ell={{I}\over{\hbar}},\\
&&&\\
\displaystyle
{\cal H}_0={{H_0}\over{\hbar\omega}},&
\displaystyle
{\cal H}_{int}={{H_{int}}\over{\hbar\omega}},&
\displaystyle
{\cal H}={{H}\over{\hbar\omega}},
\end{array}
\right\}
\label{three.nineteen}
\ee
and the dimensionless parameters,
\be
\epsilon={{\varepsilon k}\over{m\omega^2}},\quad \eta^2={{\hbar
k^2}\over{2 m\omega}},\quad \mu={{\Omega}\over{\omega}},\quad
\delta=N-\mu .
\label{three.twenty}
\ee
The parameter $\eta$ is the Lamb-Dicke parameter used in
the theory of ion traps to quantify the strength of
confinement.  It is related to effective Planck constant 
by the formula $\bar{\hbar}=2\eta^{2}$. 

\subsection{Isolated Nonlinear Resonance}
Using (\ref{three.nineteen}) and (\ref{three.twenty}), we have from
(\ref{three.fifteen}) and (\ref{three.sixteen}) the approximate
dimensionless equations of motion in the vicinity of the resonance
(\ref{three.four}),
\bea
{{d\ell}\over{d\tau}}&=&-{{\partial {\cal H}_0}\over{\partial \varphi}}=
{{\epsilon}\over{2 \eta^2}}J_N(z)\sin(\varphi+\pi
N/2),\label{three.twentyone}
\\
{{d\varphi}\over{d\tau}}&=&{{\partial {\cal H}_0}\over{\partial
\ell}}=\delta+\frac{\epsilon}{2\eta}\sqrt{{{N}\over{\ell}}}J_N^\prime(z)\cos(
\varphi+\pi
N/2),
\label{three.twentytwo}
\eea
where $z=2 \eta \sqrt{N\ell}$ and the dimensionless resonant Hamiltonian is,
\be
{\cal H}_0=\ell\delta+{{\epsilon}\over{2\eta^2}}J_N(z)\cos(\varphi+\pi
N/2).
\label{three.twentythree}
\ee
The classical dynamics corresponding the Hamiltonian
(\ref{three.twentythree}) we
shall call ``Nonlinear Resonance'' (NR).

To estimate the region of parameters of validity of Eqs
(\ref{three.twentyone}) and (\ref{three.twentytwo}), their
solutions should be compared with the solutions of exact equations
 (\ref{three.seventeen}) and (\ref{three.eighteen}).
 Equations (\ref{three.seventeen}) and (\ref{three.eighteen})
in the dimensionless variables have
the form,
\bea
{{d\ell}\over{d\tau}}&=&-\frac{\epsilon}{\eta}\sqrt{{{\ell}\over{N}}}\sin\Bigg[z
\cos\Bigg({{\varphi+\mu\tau}\over{N}}\Bigg)-\mu\tau\Bigg]\sin\Bigg
({{\varphi+\mu\tau}\over{N}}\Bigg),\label{three.twentyfour}\\
{{d\varphi}\over{d\tau}}&=&\delta-\frac{\epsilon}{2\eta}\sqrt{{{N}\over{
\ell}}}
\sin\Bigg[z
\cos\Bigg({{\varphi+\mu\tau}\over{N}}\Bigg)-\mu\tau\Bigg]\cos
\Bigg({{\varphi+\mu\tau}\over{N}}\Bigg).
\label{three.twentyfive}
\eea
Equations (\ref{three.twentyfour}) and (\ref{three.twentyfive})
are derived from the exact dimensionless Hamiltonian,
\be
{\cal
H}=\ell\delta+{{\epsilon}\over{2\eta^2}}\cos\Bigg[z\cos\Bigg({{\varphi+
\mu\tau}\over{N}}\Bigg)-\mu\tau\Bigg].
\label{three.twentysix}
\ee
In the dimensionless variables (\ref{three.nineteen}) and
(\ref{three.twenty}) , Eq. (\ref{three.three}) takes the form,
\be
{{d^2\xi}\over{d\tau^2}}+\xi=\epsilon\sin(\xi-\mu\tau).
\label{three.twentyseven}
\ee

\section{Quantum Equations of Motion}
\setcounter{equation}{0}

In the dimensionless notation (\ref{three.nineteen}) and
(\ref{three.twenty}), the quantum Hamiltonian (\ref{hfin})
takes the following form in the coordinate representation,
\be
{{\hat H}\over{\hbar\omega}}={\cal H}={{1}\over{2\eta^2}}
\Bigg[-2 \eta^{4}{{\partial^2}\over{\partial\xi^2}}+
{{\xi^2}\over{2}}+\epsilon\cos(\xi-\mu\tau)\Bigg].
\label{four.one}
\ee
The Schr\"odinger equation for the Hamiltonian (\ref{four.one}) is,
\be
i2\eta^2{{\partial\Psi(\xi,\tau)}\over{\partial\tau}}=\Bigg[
{\hat{\cal H}}_{LO}+\epsilon\cos(\xi-\mu\tau)\Bigg]
\Psi(\xi,\tau),
\label{four.two}
\ee
where ${\hat{\cal H}}_{LO}$ is the Hamiltonian of a linear oscillator,
\be
{\hat{\cal H}}_{LO}=-2\eta^4{{\partial^2}
\over{\partial\xi^2}}+{{\xi^2}\over{2}}. \nn
\ee
For $\hat h_0$ we have the well-known eigenvalue problem,
\be
{\hat{\cal H}}_{LO}|n\rangle=2\eta^2(n+1/2)|n\rangle,
\label{four.three}
\ee
where,
\be
|n\rangle\equiv\phi_n(\xi)=
\left[\frac{1}{\eta 2^n n! \sqrt{2\pi}}\right]^{1/2}
H_n(\xi/\sqrt{2}\eta)e^{-\xi^2/4\eta^2},
\label{four.four}
\ee
where $H_n(y)$ is a Hermite polynomial.  Although these eigenfunction
may appear somewhat unfamiliar because of the use of dimensionless
variables, they are in fact the standard eigenfunctions of an
unperturbed harmonic oscillator (i.e. the Hamiltonian given by
Eq. (2.1) with $\hat{H}_{I}=0$).
The normalization condition for the eigenfunction $\phi_n(\xi)$ is,
\be
\int_{-\infty}^{\infty}\phi_n(\xi)\phi_m(\xi)d\xi=\delta_{n,m} .
\ee

To describe the quantum dynamics we represent the wave
 function in (\ref{four.two}) in the form,
\be
\Psi(\xi,\tau)=\sum_{n=0}^\infty c_n(\tau)\phi_n(\xi) .
\label{four.five}
\ee
>From (\ref{four.two}) we have the equations for the complex
amplitudes,
$c_n(\tau)$,
\bea
i{{dc_m(\tau)}\over{d\tau}}&=&(m+1/2)c_m(\tau)+\frac{\epsilon}{2 \eta^{2}}
\sum_{n=0}^\infty\langle m|\cos(\xi-\mu\tau)|n\rangle c_n(\tau) \nn
\\
&=&(m+1/2)c_m(\tau)+{{\epsilon}\over{4 \eta^{2}}}\sum_{n=0}^\infty
 \Bigg(e^{-i\mu\tau}F_{m,n}(\eta)+e^{i\mu\tau}F^*_{m,n}(\eta)\Bigg)c_n(\tau).
\label{four.six}
\eea
In Eqs (\ref{four.six}), $F_{m,n}(\eta)$ is the matrix element,
\be
F_{m,n}(\eta)=\langle
m|e^{i\xi}|n\rangle={{1}\over{\sqrt{\pi2^{m+n}m!n!}}}\int^
 \infty_{-\infty}H_m(u)H_n(u)e^{-u^2+i2\eta u}du.
\label{four.seven}
\ee
Equations (\ref{four.six}) are used below, in Sec. \ref{numres}, for
numerical calculation of the
quantum dynamics of the system.

\section{Results of Numerical Calculations}\label{numres}
In this section, we present results of numerical simulations
of classical and quantum dynamics of systems whose Hamiltonians
have the form given by Eq. (\ref{hfin}). We have used
a set of parameters which will allow easy experimental
verification of our predictions using the type of
ion trap apparatus currently being used to investigate
quantum computation.

If we use the geometry for the pump and Stokes lasers
shown in Fig.1, the parameter $k$ defined by Eq. (\ref{ohkay})
is given by $k=\cos\theta (k_{p}+k_{s})\approx 2k_{0}\cos\theta$.
The laser field strengths $\left|E^{(p)}\right|$ and
$\left|E^{(s)}\right|$ can be related to the power in
the pump and the Stokes beams respectively.  It is usual
to generate one of these beams (the Stokes, say) by frequency modulation of
the pump beam, so that the beam parameters will be similar for
them both.  The power in the pump beam is given by
the formula (Ref. \cite{MilonniEberly}, p. 488)
\be
P=\frac{c\epsilon_{0}\pi}{4}w^{2}_{0}\left|E^{(p)}\right|^{2}
\ee
where $w_{0}$ is the laser spot size (i.e. the $1/e^{2}$
radius of the intensity distribution).  If we substitute this
result into Eq. (\ref{dimfuleps}), we obtain the following
expression for the dimensionless parameter $\epsilon$:
\bea
\epsilon&=&{{\varepsilon k}\over{m\omega^2}} \nn \\
&=& \frac{8A \theta P}{c k_{0} m \omega^{2}
w_{0}^{2}\Delta} s \cos^{2} , \label{trueps}
\eea
where $s = \left|E^{(s)}\right|/\left|E^{(p)}\right| $
is a dimensionless parameter of order unity which can be
controlled experimentally.

Singly ionized calcium ($m=6.64\times 10^{-26} {\rm kg}$)
is common ion in use by several groups worldwide
for ion-trap quantum research (see Fig. 3, for energy levels of this ion).  For the $^{2}S_{1/2} - ^{2}P_{1/2}$
transition in this ion (wavelength $\lambda_{0}=397\mbox{nm}$) the
Einstein A coefficient is \cite{dfvj} $A = 1.30\times 10^{8} \mbox{s}^{-1}$ and
the wavenumber is
$k_{0}=2\pi/\lambda_{0} = 1.58 \times 10^{7} \mbox{m}^{-1}$.
If we assume a
laser power of 10mW, a spot size $w_{0}=20\mu{\rm m}$
a trapping frequency $\omega = 2\pi\times 500 {\rm kHz}$, and a detuning
$\Delta=2\pi\times 1.0 {\rm GHz}$, then $\epsilon = 1333 s
\cos^{2}\theta$.  Thus by varying the experimental free parameters $s$ and
$\theta$
one can achieve a large dynamic range for the dimensionless driving
force $\epsilon$.   The Lamb-Dicke parameter $\eta$ for these
parameters is $0.502 \cos\theta$.

The driving frequency $\Omega$ is the difference between the
pump and the Stokes frequencies.  As mentioned above, these
two beams will be realized by splitting one parent beam using
a beam splitter and then frequency modulating one of the
resultant beam using either an acousto-optic or an electro-optic
modulator.  In this manner splittings as high as $\Omega =
2\pi\times 100 {\rm MHz}$ can be achieved without too much
difficulty, so that the dimensionless parameter $N$ can be in the
range 0 to 200.

\subsection{Simulation of classical dynamics}
In Figs 4a and 4b, the classical phase space (Poincare section) is
shown in $(\xi,d\xi/d\tau)$ plane, for seven initial conditions, and
for different values of the dimensionless driving force, $\epsilon$, which
characterizes the intensity of the laser beams, and is defined in
(\ref{trueps}). To derive these results, the Eq.
(\ref{three.twentyseven}) was solved numerically (for $N=4$ and
$\delta=10^{-2}$). One can see from Fig. 4a, that for small
values of $\epsilon$ ($\epsilon<2$) the classical dynamics is regular
in some regions of the phase space. Note, that even at these values
of $\epsilon$ there exist relatively large regions in the phase space
with the chaotic component. When the interaction parameter $\epsilon$
increases, the regions with the regular classical dynamics become
smaller. As one can see from Fig. 4a, even at $\epsilon=4$, the
dynamics in the region of the phase space corresponding to the
``classical ground state'' (CGS) (the vicinity of the point $(0,0))$
remains regular. At larger values of $\epsilon$ ($\epsilon\approx 8$)
the CGS becomes chaotic. This transition is demonstrated in Fig. 4b.
Figs 5 and 6, show the time evolution of the dimensionless classical
coordinate of the ion, $\xi(\tau)$. The maximum dimensionless time of
simulations, $\tau_{max}=100$, corresponds to the real time-scale
$t_{max}= 31.8 \mu {\rm s}$ ($\omega = 2\pi \times 500 {\rm kHz}$). The
initial
conditions for the dynamics shown in Figs 5 and 6 correspond to the
CGS. As one can see from Fig. 5, the chaotic component appears at
$\epsilon\approx 8$, and is well resolved for $\epsilon>10$ (see Fig.
6). The frequency Fourier transform for the dynamics shown in Figs 5
and 6, is presented in Figs 7 and 8. It is well-known, that the
transition to the dynamical chaos in classical dynamical systems is
accompanied by the transition in the frequency Fourier spectrum. The
regular dynamics corresponds to the discrete frequency spectrum, the
chaotic dynamics corresponds to the continuous frequency spectrum.
This characteristic modification of the frequency spectrum is
demonstrated in Figs 7 and 8. For small values of the dimensionless
driving force $\epsilon$, one can see only some quasi-discrete lines in
the Fourier spectrum. When $\epsilon$ increases, the frequency
spectrum transforms to the continuous one.

\subsection{Simulation of quantum dynamics}
To simulate a quantum dynamics the following parameters were chosen:
$N=4$, $\delta=10^{-2}$, $\eta=0.45$. For the initial
conditions we used the ground state of the unperturbed quantum linear
oscillator: $c_0(0)=1$, $c_n(0)=0$ for $n>0$. Fig. 9, shows the time
evolution of the quantum probabilities
$P_n(\tau)\equiv |c_n(\tau)|^2$ (n=0, 1, 2, 3, 4 and 5) for relatively small
values of $\epsilon=$0, 0.5, 1, 1.5 and 2. These values of $\epsilon$
correspond to the regular classical dynamics which starts from the
CGS. Time evolution of quantum probabilities $P_n(\tau)$
($n=$ 0, 1, 2 and 3) for bigger values of $\epsilon$ is shown in Fig. 10.
Because the value of the parameter $\epsilon\approx 8$ corresponds to
the classical chaotic dynamics for the initially populated CGS, the
curves (4) in Fig. 10 describe the quantum chaotic motion. Fig. 11,
shows the dynamics of the probability function $P_0(\tau)$ for the
larger time interval: $\tau\in[0,30]$. In the real time this
correspond to: $t\in[0, 9.54] \mu{\rm s}$.
As one can see from Fig. 11, for $\epsilon>7.5$ the
dynamics of the probability function $P_0(\tau)$ becomes rather
complicated, and correspond to the classical chaotic motion. We hope
that this complicated dynamics of the probability functions
$P_n(\tau)$ can be measured directly in the experiments with trapped
ion. Fig. 12, represents the results of the numerical simulations of
the frequency Fourier spectrum, $P_0(\nu)$, of the quantum
probability function $P_0(\tau)$. As one can see from Fig. 12, the
characteristic qualitative modification of the frequency spectrum
$P_0(\nu)$ starts from $\epsilon>7.5$. This modification of the
frequency Fourier spectrum can also be measured experimentally. In
Fig. 13, we show the results of numerical simulation
of the dynamical evolution of the average value:
$\langle\xi^2\rangle\equiv\langle\Psi(\xi,\tau)|\xi^2|\Psi(\xi,\tau)\rangle$,
where the wave function $\Psi(\xi,\tau)$ is defined in (4.7). This
dynamical characteristic is very important for understanding the
conditions of stability of the system under consideration, as it
describes the amplitude of the ion's oscillations in a trap due to
the influence of the resonant laser fields. Experimental measuring
the time dependence of this amplitude is important for characterizing
the regular and chaotic dynamics of an ion. The frequency spectrum of
the amplitude $\langle\xi^2\rangle$ is shown in Fig. 14 for different
values of the perturbation parameter $\epsilon$. As one can see from
Fig. 14, at $\epsilon>7.5$, the frequency spectrum qualitatively
modifies, and includes many harmonics. In Fig. 15, we show the time
evolution of two functions:
the unperturbed Hamiltonian of quantum linear oscillator (4.3),
$\langle {\hat{\cal H}}_{LO}\rangle$, and the amplitude $\langle\xi^2\rangle$,
for the values of the perturbation parameter $\epsilon= 0; 05; 2$.
The corresponding classical dynamics is regular in this case. The
time evolution of the same functions is shown in Fig. 16 for larger
values of the parameter $\epsilon=3;5;7.5; 8$. The curves (3) and (4)
correspond to the classically chaotic regime of motion. The maximum
simulation time at Fig. 16 is: $t_{max}=4.77 \mu{\rm s}$.

\section{Conclusion}
\setcounter{equation}{0}
In this paper, we introduced a quantum model which describes a
transition from a regular dynamics to quantum chaos, for a single ion
in a linear ion trap. The configuration of the resonant laser fields
allowed us to represent the model in a ``standard'' form. Namely, our
model formally describes a quantum linear oscillator interacting with
a one-dimensional plane wave (see the Hamiltonian (2.29)). The
classical version of this model is very important and useful in the
theory of the dynamical chaos \cite{zas1}. This model differs
significantly from the ``usual'' nonlinear models (as, for example, a
``standard map''\cite{chir2,ll}) considered in the theoretical works
on dynamical chaos. Namely, the model described by the Hamiltonian
(2.19) includes ``nonlinearity'' and ``perturbation'' in the same
term. In this sense, this model is degenerated, and possesses many
unusual properties \cite{zas1}. So, quantum analysis of this model
both theoretically and experimentally will be of significant
importance for understanding complicated dynamics in this system, and
for applications in different devices based on the trapped ions,
including quantum computer. The numerical results presented in this
paper describe both regular and chaotic quantum dynamics which starts
from the initial population of the linear oscillator's ground state.
We believe that this regime of motion should be investigated in the
experiments with the trapped single ion. These experiments will allow
one to establish qualitative and quantitative correspondence between
the quantum dynamics described by the Hamiltonian (2.29) and the real
system ``a single trapped ion + resonant laser fields''. Further
theoretical analysis and experiments should include both pure quantum
and quasiclassical regions of initial population, and the parameters
describing regular and chaotic regimes in both these regions. These
investigations are now in progress.

\section*{Acknowledgments}
It is a pleasure to thank G.D. Doolen, G. Milburn, and J. Rehacek 
for valuable discussions. This work was
partly supported by the National Security Agency, and by the
Department of Energy under contract W-7405-ENG-36.

\section*{Appendix A: Derivation of Eq. (\ref{aches})}
\renewcommand{\theequation}{{\rm
A.\arabic{equation}}}\setcounter{equation}{0}

The dipole matrix elements appearing in Eq. (\ref{hthree})
can be written in terms of the Einstein A coefficient
between the upper and lower manifolds as follows:
\be
\bra{\mu}\hat{d}_{i}\ket{\lambda} =
\sqrt{\frac{3 A c^{2}\left( 2J_{\lambda}+1\right)}{4 \om{0}^{2}
\alpha}}
\sum_{q=-1}^{1}
\threej{J_{\mu}}{1}{J_{\lambda}}{-m_{\mu}}{q}{m_{\lambda}}e^{(q)}_{i}
\label{dme}
\ee
where $c$ is the speed of light, $\alpha$ is the fine structure
constant, $\om{0}$ is the angular frequency of the transition
between the upper and lower manifolds, $(J_{\lambda}, m_{\lambda})$
are the magnetic quantum numbers for the upper manifold state
$\ket{\lambda}$
$(J_{\mu}, m_{\mu})$
are the magnetic quantum numbers for the lower manifold state
$\ket{\mu}$,
the term contain six quantities in brackets is the Wigner 3-j symbol,
and $e^{(q)}_{i}$ is the $i$-th component of the $q$-th normalized
spherical basis vector, viz.
\bea
{\bf e}^{(1)}&=&-\frac{1}{\sqrt{2}}(1,-i,0) , \\
{\bf e}^{(0)}&=&(0,0,1) , \\
{\bf e}^{(-1)}&=&\frac{1}{\sqrt{2}}(1,i,0) . \\
\eea
Substituting Eq. (\ref{dme}) into Eq. (\ref{hthree}), and assuming
that
the Zeeman splitting in the upper manifold is small compared to
the overall detuning, so that $\Delta_{\lambda}\approx \Delta$
(independent of $\lambda$), then we find that the quantities
${\cal h}_{\mu,\nu}$ are given by the following formula:
\be
\kappa_{\mu,\nu}\left(\hat{\bf{r}},t\right)=
\frac{3\pi \epsilon_{0} A c^{3}}{4 \om{0}^{3} \Delta}
\Lambda_{ij}\left(\mu,\nu\right)
E_{i}\left(\hat{\bf{r}},t\right)
E^{\ast}_{j}\left(\hat{\bf{r}},t\right),
\ee
where the tensor $ \Lambda_{ij}\left(\mu,\nu\right) $
is given by
\be
\Lambda_{ij}\left(\mu,\nu\right)=
\left( 2J_{\lambda}+1\right)
\sum_{m_{\lambda}=-J_{\lambda}}^{J_{\lambda}}
\sum_{q,q'=-1}^{1}
\threej{J_{\mu}}{1}{J_{\lambda}}{-m_{\mu}}{q}{m_{\lambda}}
\threej{J_{\nu}}{1}{J_{\lambda}}{-m_{\nu}}{q'}{m_{\lambda}}
e^{(q)}_{i}
e^{(q')\ast}_{j}
\ee

If we assume that the lower manifold is
the $^{2}S_{1/2}$ ground state of an alkali ion,
the two states being denoted $\ket{1}= ^{2}S_{1/2}, m=-1/2$
and $\ket{2}= ^{2}S_{1/2}, m=1/2$, and that the
upper manifold is the $^{2}P_{1/2}$ state,
then these tensors can be found in closed form:
\bea
\Lambda_{ij}\left(1,1\right)&=&\Lambda_{ij}^{\ast}\left(2,2\right)=
\frac{1}{3}
\left(\begin{array}{ccc}
1&-\ci&0\\
\ci&1&0\\
0&0&1
\end{array}\right),\nn \\
\Lambda_{ij}\left(1,2\right)&=&-\Lambda_{ij}^{\ast}\left(2,1\right)=
\frac{1}{3}
\left(\begin{array}{ccc}
0&0&-1\\
0&0&-\ci\\
1&\ci&0
\end{array}\right).
\eea
Therefore the cross products involving the electric field
components with these tensors can be written as follows:
\bea
\Lambda_{ij}\left(1,1\right)E_{i}E^{\ast}_{j}&=&\frac{1}{3}
\left[\left|{\bf E}\right|^{2}+2 {\rm Im}
\left\{E_{X}E_{Y}^{\ast}\right\}\right],\\
\Lambda_{ij}\left(2,2\right)E_{i}E^{\ast}_{j}&=&\frac{1}{3}
\left[\left|{\bf E}\right|^{2}-2 {\rm Im}
\left\{E_{X}E_{Y}^{\ast}\right\}\right],\\
\Lambda_{ij}\left(1,2\right)E_{i}E^{\ast}_{j}&=&\frac{1}{3}
\left[E_{Z}\left(E^{\ast}_{X}+\ci E^{\ast}_{Y}\right)-
E^{\ast}_{Z}\left(E_{X}+\ci E_{Y}\right)\right], \\
\Lambda_{ij}\left(2,1\right)E_{i}E^{\ast}_{j}&=&\frac{1}{3}
\left[-E_{Z}\left(E^{\ast}_{X}-\ci E^{\ast}_{Y}\right)+
E^{\ast}_{Z}\left(E_{X}-\ci E_{Y}\right) \right].
\eea
If these results are substituted into the definition of $h_{0}$
and ${h_{i}}$, Eqs (\ref{hdefstart})-(\ref{hdefend}),
one obtains Eq. (\ref{aches}).  Similar results, with slightly
different numerical factors, are obtained if the upper manifold
is a $^2P_{3/2}$ state.

\section*{Appendix B: Canonical Transform to ``action-angle''
variables}
\renewcommand{\theequation}{{\rm
B.\arabic{equation}}}\setcounter{equation}{0}

The theory of canonical transforms in classical
mechanics is described in detail in Ref. \cite{LL}, \S 45.
We want to transform from a set of variable $x,p$ to
a new set $\varphi,I$, where $\varphi$ plays the role
of position coordinate and $I$ the role of momentum.
Such transforms are specified by a {\em generating
function}, $F(x,\varphi,t)$.  Then variables $p$ and $I$
and the Hamiltonian in the new coordinate system are
are related to $F$ by the following formulas:
\bea
p&=&\frac{\partial F}{\partial x} , \\
I&=&-\frac{\partial F}{\partial \varphi}, \\
H(\varphi,I)&=&H(x,p)+\frac{\partial F}{\partial t}
\eea

For the canonical transform used in section 3.1,
the generating function is given by
\be
F(x,\varphi,t)=-\frac{m\omega}{2}x^{2}\tan\left(\frac{\varphi+\Omega
t}{N}\right) .
\ee
Substituting, we find:
\bea
p&=&-m\omega x \tan\left(\frac{\varphi+\Omega
t}{N}\right) \label{peepee}\\
I&=&-\frac{m\omega}{2N}x^{2}\sec^{2}\left(\frac{\varphi+\Omega
t}{N}\right) \label{sesad}\\
H(\varphi,I)&=&H(x,p)-\frac{m\omega\Omega}{2N}x^{2}\sec^{2}\left(\frac{\varphi+\
Omega
t}{N}\right)\\
&=& H(x,p)-\Omega I .
\eea
Equation (\ref{peepee}) implies that
\be
x=\sqrt{\frac{2NI}{m\omega}}\cos\left(\frac{\varphi+\Omega
t}{N}\right).
\ee
On substitution of this last formula into Eq. (\ref{sesad})
we obtain
\be
p=-\sqrt{2Nm\omega I}\sin\left(\frac{\varphi+\Omega
t}{N}\right).
\ee

\newpage
\noindent
{\bf Figure captions}
\vspace{0.5cm}

\noindent
{\bf Fig. 1:}  A schematic diagram of an ion in a linear trap to
illustrate the notation and configurations of the laser fields.

\vspace{0.3cm}

\noindent
{\bf Fig. 2:}  A schematic illustration of the energy levels of a
trapped ion.

\vspace{0.3cm}

\noindent
{\bf Fig. 3:}  Energy levels of Ca$^+$ ion. Wavelengths and radiative 
lifetimes are shown. See \cite{dfvj} for references.

\vspace{0.3cm}

\noindent
{\bf Fig. 4:}  Classical phase space (Poincare section). Trajectories
with different initial conditions are shown.  The values of
$\epsilon$ are indicated in the figure: (a) $\epsilon=2,2.5,3,4$; (b)
$\epsilon=5,8,10,20$; $\eta=0.45$; $N=4$, $\delta=10^{-2}$.

\vspace{0.3cm}

\noindent
{\bf Fig. 5:}  Time evolution of the dimensionless classical
amplitude of oscillations $\xi(\tau)$ for different values of
$\epsilon=0,0.5,1,2,5,8$. The maximum time of simulation it
$t_{max}=31.8\mu{\rm s}$. Initial condition: $(\xi,d\xi/d\tau)=(0,0)$.

\vspace{0.3cm}

\noindent
{\bf Fig. 6:}  Chaotic dynamics of the dimensionless classical
amplitude of oscillations $\xi(\tau)$ for the values of
$\epsilon=10,20,30,40$. The maximum time of simulation it
$t_{max}=31.8\mu{\rm s}$. Initial condition: $(\xi,d\xi/d\tau)=(0,0)$.

\vspace{0.3cm}

\noindent
{\bf Fig. 7:}  Frequency Fourier spectrum of the classical amplitude
$\xi(\tau)$ for $\epsilon$=0,0.5,1,2,5 and 8.
Transition to chaos appears at $\epsilon\approx$ 8. Initial
condition: $(\xi,d\xi/d\tau)=(0,0)$.

\vspace{0.3cm}

\noindent
{\bf Fig. 8:}  Frequency Fourier spectrum of the chaotic dynamics of
the classical amplitude $\xi(\tau)$ for $\epsilon=$10,20,30 and 40.
Transition to chaos appears at $\epsilon\approx$8. Initial
condition: $(\xi,d\xi/d\tau)=(0,0)$.

\vspace{0.3cm}

\noindent
{\bf Fig. 9:}  Dynamics of the quantum probabilities:
$P_n(\tau)=|c_n(\tau)|^2$, for $n=$0,1,2,3,4 and 5, and for $\epsilon =$
0, 0.5, 1, 1.5 and 2. Initial condition: $c_0(0)=1$, $c_n(0)=0$ for
($n>0)$.

\vspace{0.3cm}

\noindent
{\bf Fig. 10:}  Dynamics of the quantum probabilities:
$P_n(\tau)=|c_n(\tau)|^2$, for $n=$0, 1, 2 and 3, and for $\epsilon
=$3, 5 and 7.5. Initial condition: $c_0(0)=1$, $c_n(0)=0$ for ($n>0)$.

\vspace{0.3cm}

\noindent
{\bf Fig. 11:}  Dynamics of the quantum probability:
$P_0(\tau)=|c_0(\tau)|^2$, for $\epsilon =$1, 5, 7.5 and 8. Initial
condition: $c_0(0)=1$, $c_n(0)=0$ for ($n>0)$. Transition to quantum
chaos corresponds to $\epsilon\approx 8$. Initial conditions as in
Fig. 10.

\vspace{0.3cm}

\noindent
{\bf Fig. 12:}  Frequency Fourier spectrum of the quantum
probability: $P_0(\tau)=|c_0(\tau)|^2$ for $\epsilon =$1, 5, 7.5 and 8.
Transition to chaos appears at $\epsilon\approx 8$. Initial
conditions as in Fig. 10.

\vspace{0.3cm}

\noindent
{\bf Fig. 13:}  Dynamical evolution of the average value:
$\langle\xi^2\rangle\equiv\langle\Psi(\xi,\tau)|\xi^2|\Psi(\xi,\tau)\rangle$,
where the wave function $\Psi(\xi,\tau)$ is defined in (4.7).
$\epsilon =$3, 5, 7.5 and 8.
Transition to chaos appears at $\epsilon\approx 8$. Initial
conditions as in Fig. 10.

\vspace{0.3cm}

\noindent
{\bf Fig. 14:}  Frequency Fourier spectrum of the quantum amplitude:
$\langle\xi^2\rangle$, for $\epsilon =$1, 5, 7.5 and 8
Transition to chaos appears at $\epsilon\approx$ 8. Initial
conditions as in Fig. 10.

\vspace{0.3cm}

\noindent
{\bf Fig. 15:}  Time evolution of two functions:
the unperturbed Hamiltonian of quantum linear oscillator (4.3),
$\langle {\hat{\cal H}}_{LO}\rangle$, and the amplitude $\langle\xi^2\rangle$,
for the values of the perturbation parameter $\epsilon=$ 0, 0.5 and 2.
The corresponding classical dynamics is regular in this case.
$\epsilon=$0, 0.5, 2. Initial conditions as in Fig. 10.

\vspace{0.3cm}

\noindent
{\bf Fig. 16:}   The time evolution of the same functions shown in
Fig. 14 but for larger values of the parameter $\epsilon=$3, 5, 7.5 and 8.
The curves (3) and (4) correspond to the classically chaotic regime
of motion.
 Initial conditions as in Fig. 10.

\vspace{0.3cm}

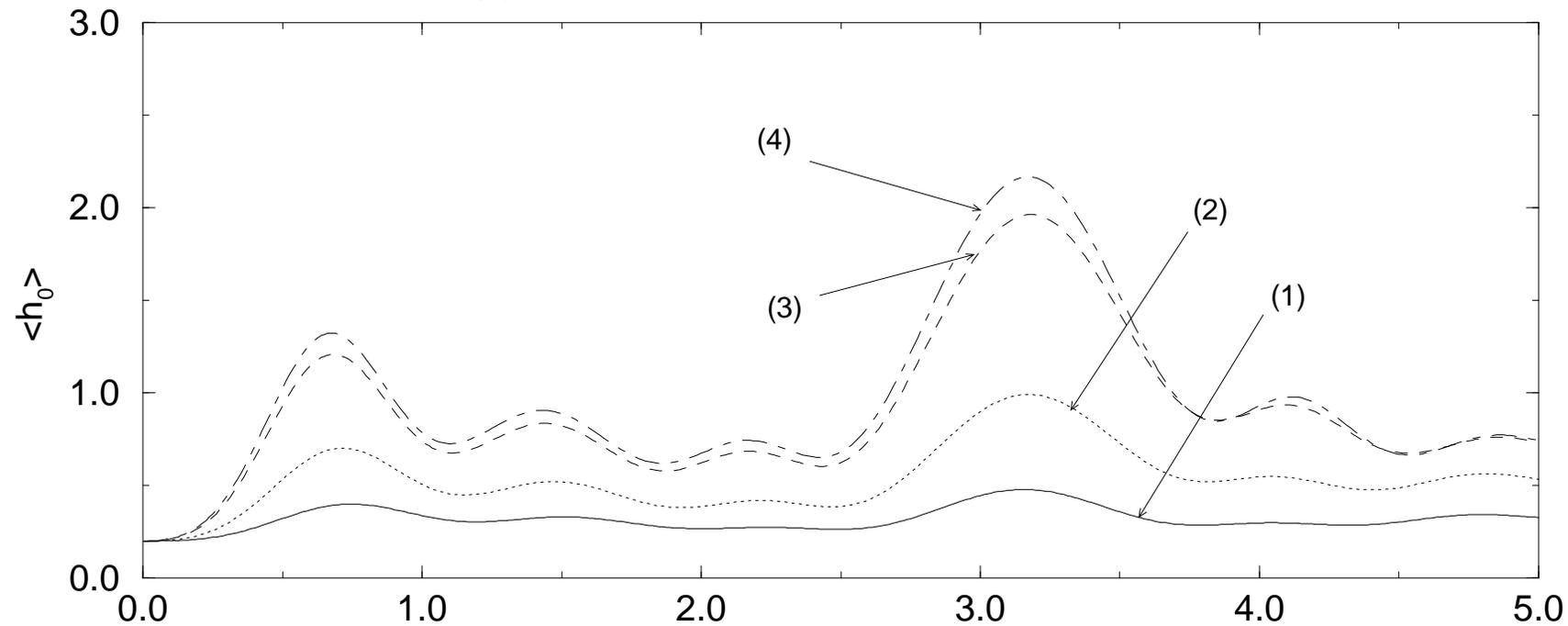
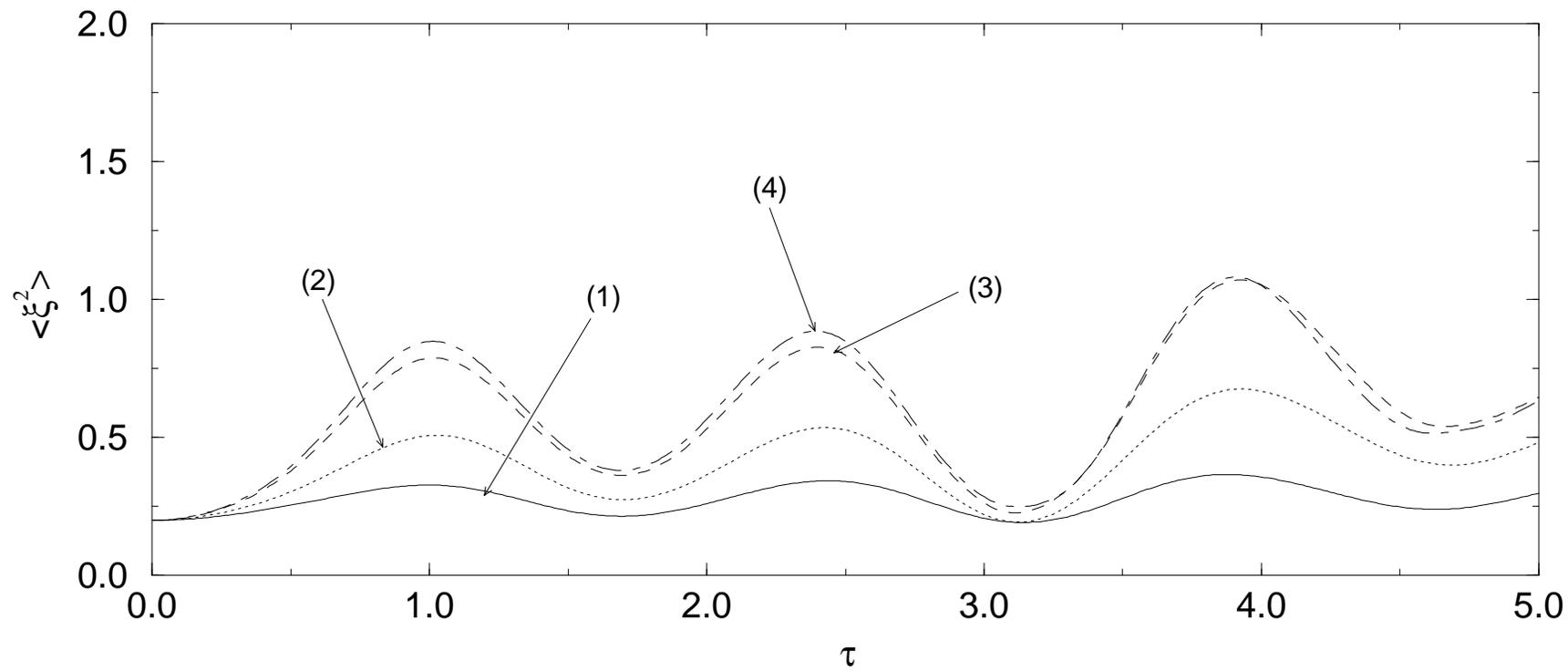
\end{document}